\documentclass[pra,aps,twocolumn,superscriptaddress,nofootinbib]{revtex4}

\begin{document}

\title{Algebraic and information-theoretic conditions for 
operator quantum error-correction}

\author{Michael A. Nielsen} 
\affiliation{School of Physical Sciences,
The University of Queensland, Queensland 4072, Australia}

\author{David Poulin} 
\affiliation{School of Physical Sciences,
The University of Queensland, Queensland 4072, Australia}

\date{\today}

\begin{abstract}
  Operator quantum error-correction is a technique for robustly
  storing quantum information in the presence of noise.  It
  generalizes the standard theory of quantum error-correction, and
  provides a unified framework for topics such as quantum
  error-correction, decoherence-free subspaces, and noiseless
  subsystems.  This paper develops (a) easily applied algebraic and
  information-theoretic conditions which characterize when operator
  quantum error-correction is feasible; (b) a representation theorem
  for a class of noise processes which can be corrected using operator
  quantum error-correction; and (c) generalizations of the coherent
  information and quantum data processing inequality to the setting of
  operator quantum error-correction.
\end{abstract}


\maketitle

To develop quantum technologies such as quantum computers and quantum
communication networks, it will be necessary to protect quantum
systems against the effects of noise.  Considerable progress toward
this goal was made in the late 1990s, when a theory of fault-tolerant
quantum computing was
developed~\cite{Shor96b,Kitaev97a,Knill98a,Aharonov97a,Preskill98b},
based on the theory of quantum error-correcting
codes~\cite{Shor95a,Steane96c,Knill97a,Gottesman96a,Calderbank97b}.

The early theory of quantum error-correcting codes was based on the
following ideas: (1) quantum information is stored in a subspace $A$
of a larger state space $V = A \oplus C$.  $A$ is known as the
\emph{code space}, while $V$ is the state space of the physical system
being used to store the information; (2) some physically-motivated
noise process corrupts the physical system; (3) a recovery step is
performed, restoring the original quantum information stored in $A$.

Since its development this theory has been refined and generalized in
a variety of ways, notably through the introduction of
decoherence-free
subspaces~\cite{Palma96a,Duan97a,Zanardi97a,Lidar98a}, noise-free
subsystems~\cite{Knill00b,Zanardi01a,Kempe01b}, and of operator
quantum error-correction.  In particular, the framework of
\emph{operator quantum error-correction}~\cite{Kribs05a,Kribs05b}
provides a single framework integrating and unifying all of these
techniques.

Operator quantum error-correction is based on the following ideas: (1)
quantum information is stored in a space $A$ which appears as a tensor
factor in a subspace of the overall state space, $V$, i.e., $V = (A
\otimes B) \oplus C$; (2) some physically-motivated noise process
corrupts the physical system; (3) a recovery step is performed,
restoring the original encoded quantum information stored in $A$.

Operator quantum error-correction is a significant generalization of
standard quantum error-corection.  Kribs \emph{et
  al}~\cite{Kribs05a,Kribs05b} have shown that operator quantum
error-correction provides a natural framework unifying and
generalizing earlier approaches, including standard quantum
error-correction, decoherence-free subspaces, and noiseless
subsystems.  Bacon~\cite{Bacon05a} has recently exhibited interesting
examples in which operator quantum error-correction plays a critical
role.

The purpose of this paper is to develop easily applied necessary and
sufficient conditions for operator quantum error-correction.  In
particular, we obtain a set of algebraic conditions characterizing
operator quantum error-correction.  These conditions generalize the
well-known conditions for standard quantum
error-correction~\cite{Bennett96a,Knill97a}, which are the basis for
the theory of quantum error-correcting codes, enabling the
construction of large classes of
codes~\cite{Gottesman96a,Calderbank97a}.  The necessity of these
conditions for operator quantum error-correction was proved
in~\cite{Kribs05a}, but the proof of sufficiency was left open.  We
establish the sufficiency of these conditions, and use the conditions
to establish an elegant representation theorem for a class of noise
processes which can be corrected using operator quantum
error-correction.

We also prove a set of information-theoretic conditions characterizing
operator quantum error-correction, based on generalizations of the
coherent information and the quantum data processing inequality.  In
the context of quantum error-correction codes these concepts were
developed in~\cite{Schumacher96b}, and were critical in developing the
theory of quantum channel
capacity~\cite{Schumacher96a,Schumacher96b,Barnum98a,Barnum98b,Lloyd97a,Shor02a,Devetak05a}.

\textbf{Definition of operator quantum error-correction:} Suppose $V$
is the Hilbert space for some quantum system, and we decompose $V = (A
\otimes B) \oplus C$ for some choice of $A,B$ and $C$.  Suppose ${\cal
  E}$ is a quantum operation acting on $V$.  Then we say $A$ is an
${\cal E}$-correcting subsystem with respect to the decomposition $V =
(A \otimes B) \oplus C$ if there exists a trace-preserving quantum
operation ${\cal R}$ (the recovery operation) such that for all $\rho$
with support on $A$, and all $\sigma$ with support on $B$, we have
$({\cal R} \circ {\cal E})(\rho \otimes \sigma) \propto \rho \otimes
\sigma'$, for some $\sigma'$ with support on $B$.  Physically, this
means that we can store information in the subsystem $A$, and recover
the information after noise ${\cal E}$ by applying the recovery
operation ${\cal E}$.  Quantum error-correcting codes arise as the
special case of this definition where $B$ is trivial (i.e.,
one-dimensional), which is equivalent to decomposing $V = A \oplus C$.
That is, in an error-correcting code we encode information in a
\emph{subspace}, while in an error-correcting subsystem we may encode
information in a \emph{subsystem} of a subspace.


\textbf{Algebraic characterization of operator quantum
  error-correction:} Suppose ${\cal E}(\rho) = \sum_j E_j \rho
E_j^\dagger$ is an operator-sum representation for ${\cal E}$ in terms
of operation elements $E_j$.  We will prove that the following two
conditions are equivalent:

\textbf{[a]}: $A$ is an ${\cal E}$-correcting subsystem with respect
to the decomposition $V = (A \otimes B) \oplus C$.

\textbf{[b]}: $PE_j^\dagger E_k P = I_A \otimes B_{jk}$ for all $j$
and $k$, where $P$ projects onto $A\otimes B$, and the $B_{jk}$ are
operators on $B$.

Condition~\textbf{[b]} provides an elegant and easily-checkable set of
necessary and sufficient conditions for operator quantum
error-correction, generalizing the standard quantum error-correction
conditions~\cite{Bennett96a,Knill97a}.

\textbf{Proof that \textbf{[a]} implies \textbf{[b]}:} This was proved
in~\cite{Kribs05a}, and is a straightforward generalization of the
corresponding part of the proof of the quantum error-correction
conditions as given in, e.g., Chap.~10 of~\cite{Nielsen00a}.  One of
the ideas used in the proof is used again later, so for completeness
we give a brief outline.  Suppose the recovery operation ${\cal R}$
has operation elements $R_j$.  Define an operation ${\cal P}(\rho)
\equiv P \rho P$.  Then it can be shown that ${\cal R} \circ {\cal E}
\circ {\cal P} = {\cal I}_A \otimes {\cal N}$ for some operation
${\cal N}$ on system $B$.  Standard results (see, e.g.,~Chap.~9
of~\cite{Nielsen00a}) about the unitary freedom in operation elements
imply that $R_j E_k P = I \otimes N_{jk}$ for some set of operators
$N_{jk}$ acting on system $B$.  Multiplying this equation by its
adjoint, for a suitable choice of indices we obtain $P E_l^\dagger
R_j^\dagger R_j E_k P = I \otimes N_{jl}^\dagger N_{jk}$.  Summing
over $j$ and using the fact that ${\cal R}$ is trace-preserving (i.e.,
$\sum_j R_j^\dagger R_j = I$) gives the result. \textbf{QED}

We will give two proofs that \textbf{[b]} implies \textbf{[a]}.  The
first proof is deeper, and is based on a third equivalent condition,
\textbf{[c]}; we prove \textbf{[b]} $\Rightarrow$ \textbf{[c]}
$\Rightarrow$ \textbf{[a]}.  \textbf{[c]} has many rich consequences,
including the information-theoretic characterization of operator
error-correction described later, and a beautiful representation
theorem (described below) for correctable ${\cal E}$ in the special
case when $V = A \otimes B$.  Our second proof that \textbf{[b]}
implies \textbf{[a]} is a more straightforward extension of the
standard quantum error-correction conditions.  This proof is arguably
simpler than the first, but does not appear to have the same rich
consequences, and so we merely provide a sketch.

To state condition \textbf{[c]} involves a somewhat elaborate
construction involving auxiliary systems, inspired
by~\cite{Schumacher96b}.  We introduce systems $R_A$ and $R_B$ whose
Hilbert spaces are copies of $A$ and $B$, respectively.  We define
(unnormalized) maximally entangled states $|\alpha\rangle \equiv
\sum_j |j\rangle |j\rangle$ of $R_A A$ and $|\beta\rangle \equiv
\sum_k |k\rangle |k\rangle$ of $R_B B$.  The state $|\alpha\rangle
|\beta\rangle$ may be regarded as a joint state of $R_A R_B V$ in a
natural way.

Next, we introduce a system $E$ which will act as a model environment
for the operation ${\cal E}$.  We suppose $E$ has an orthonormal basis
$|j\rangle$ whose elements are in one-to-one correspondence with the
operation elements $E_j$.  Supposing $|s\rangle$ is some fixed initial
state of $E$, we define a linear operation $L$ on $V E$ which has the
action $L|\psi\rangle |s\rangle \equiv \sum_j E_j |\psi\rangle
|j\rangle$.  Note that the effect of $L$ on $V E$, after tracing out,
is equivalent to the action of ${\cal E}$ on $V$.

Define a state $|\psi'\rangle \equiv (I_{R_A R_B} \otimes L)
|\alpha\rangle |\beta\rangle |s\rangle$.  $|\psi'\rangle$ can be
thought of as the combined state of $R_A R_B V E$ \emph{after} the
noise is applied.  We define a corresponding density matrix $\rho'
\equiv |\psi'\rangle \langle \psi'|$, and use notations like
$\rho_{R_BE}'$ to denote the result when all systems but $R_B$ and $E$
are traced out.  With these definitions we may state
condition~\textbf{[c]}.

\textbf{[c]}: $\rho_{R_A R_B E}' = \rho_{R_A}' \otimes \rho_{R_B E}'$.

\textbf{Proof that [b] implies [c]:} The definition of $\rho'$ and a
direct calculation shows that:
\begin{eqnarray} \label{eq:reference_identity}
  \rho_{R_A R_B E}' = \sum_{jk} PE_j^T E_k^* P \otimes |j\rangle \langle k|,
\end{eqnarray}
where $PE_j^TE_k^*P$ is understood as an operator on $R_AR_B$.  To do
this we identify the bases $|j\rangle_{R_A}$ and $|j\rangle_A$, and
take the complex conjugate and transpose with respect to this basis.
Taking the complex conjugate of \textbf{[b]} and substituting gives
the desired result.  (The converse, that \textbf{[c]} implies
\textbf{[b]}, also follows directly from
Eq.~(\ref{eq:reference_identity}), although we will not need this
implication.) \textbf{QED}

\textbf{Proof that [c] implies [a]:} (c.f.~\cite{Schumacher96b}) We
Schmidt decompose $|\psi'\rangle$ with respect to the bipartite
decomposition $R_A R_B E : V$.  Making use of the fact that the
Schmidt vectors of $R_A R_B E$ are eigenvectors of $\rho_{R_A R_B E}'
= \rho_{R_A}' \otimes \rho_{R_B E}'$, this gives rise to the Schmidt
form (this and subsequent states are only written up to
normalization):
\begin{eqnarray}
  |\psi'\rangle = \sum_{jk} \sqrt{q_k} |j\rangle_{R_A} |k\rangle_{R_B E}
  |e_{jk}\rangle_V,
\end{eqnarray}
where the $|j\rangle_{R_A}$ are orthonormal eigenvectors of
$\rho_{R_A}'$, the $|k\rangle_{R_B E}$ and $q_k$ are orthonormal
eigenvectors and eigenvalues of $\rho_{R_B E}'$, and the
$|e_{jk}\rangle_V$ are orthonormal Schmidt vectors on $V$.

Define an orthonormal set of projectors $P_k \equiv \sum_j
|e_{jk}\rangle_V \langle e_{jk}|$ acting on $V$. We define the first
step of recovery ${\cal R}$ to be performing a measurement of $P_k$,
resulting in the state:
\begin{eqnarray}
  |\psi'_k\rangle = \sum_{j} |j\rangle_{R_A} |k\rangle_{R_B E}
  |e_{jk}\rangle_V.
\end{eqnarray}
The second and final step of recovery is to apply a unitary $U_k$
which takes $|e_{jk}\rangle_V$ to $|j\rangle_A |s\rangle_B$, where
$|s\rangle_B$ is some standard state of $B$.  The net effect of the
recovery procedure is to produce the following state of $R_A R_B V E$:
\begin{eqnarray}
  |\psi''_k\rangle = \sum_{j} |j\rangle_{R_A} |j\rangle_A |s\rangle_B 
  |k\rangle_{R_B E}
\end{eqnarray}
Thus, we have restored the initial maximal entanglement between $R_A$
and $A$.  

Summarizing, we have shown that if $R_A A$ and $R_B B$ each start out
maximally entangled, and we apply the noise ${\cal E}$ followed by the
recovery ${\cal R}$ to $V$, then the resulting state of $R_A A$ is the
original maximally entangled state.  Standard techniques
(e.g.,~\cite{Schumacher96a}) imply that we must have $({\cal R} \circ
{\cal E})(\rho \otimes \sigma) = \rho \otimes \sigma'$ for all $\rho$
on system $A$ and all $\sigma$ on system $B$. \textbf{QED}


\textbf{Representation theorem for correctable operations:} When $V =
A \otimes B$, i.e., when $C$ is trivial, the proof that \textbf{[c]}
implies \textbf{[a]} has as a consequence the elegant representation
${\cal E} = {\cal U} \circ ({\cal I}_A \otimes {\cal N}_B)$ for some
noisy operation ${\cal N}_B$ on $B$ alone, and some unitary operation
${\cal U}$ on $V$.

To see this, note that when $V = A \otimes B$ the recovery procedure
may be modified, omitting the step where $P_k$ is measured, and
instead simply applying a single unitary operation $W |e_{jk}\rangle_V
\equiv |j\rangle_A |k\rangle_B$.  If ${\cal W}$ is the quantum
operation corresponding to $W$ then we see that ${\cal W} \circ {\cal
  E} = {\cal I}_A \otimes {\cal N}_B$, so using ${\cal U} \equiv {\cal
  W^\dagger}$ gives the desired representation. \textbf{QED}

\textbf{Alternate proof that [b] implies [a] (sketch):} Fix a state
$\sigma = |s\rangle \langle s|$ of $B$, and define a quantum operation
${\cal E}_s(\rho) \equiv {\cal E}(\rho \otimes \sigma)$ mapping states
of $A$ to states of $V$.  We will use condition \textbf{[b]} to show
that there exists a \emph{single} universal recovery operation ${\cal
  R}$ which acts as a recovery operation for \emph{all} ${\cal E}_s$.
Linearity then implies that $({\cal R} \circ {\cal E})(\rho \otimes
\sigma) = \rho \otimes \sigma'$ for all $\rho$ and $\sigma$.

To prove this, note that a set of operation elements for ${\cal E}_s$
is the set $E_{j,s} : A \rightarrow V$ defined by $E_{j,s} \equiv E_j
P |s\rangle$.  That is, ${\cal E}_s(\rho) = \sum_j E_{j,s} \rho
E_{j,s}^\dagger$.  (This can be verified by a calculation.)  We will
show that the set of errors $E_{j,s}$, where $j$ and $|s\rangle$ are
\emph{both} allowed to vary over all possible values, is a correctable
set of errors mapping $A$ to $V$, in the sense of standard
error-correction.  This suffices to establish the existence of a
single universal recovery operation ${\cal R}$ which acts as a
recovery operation for all ${\cal E}_s$.  To see this, note that using
\textbf{[b]} we obtain
\begin{eqnarray}
  I_A E_{j,s}^\dagger E_{k,t}I_A = \langle s| P E_j^\dagger E_k P|t\rangle
  = e_{jkst} I_A,
\end{eqnarray}
for complex numbers $e_{jkst}$.  Thus the standard error-correction
conditions apply, which suffices to establish the existence of a
suitable recovery ${\cal R}$. \textbf{QED}

\textbf{Linearity of the set of correctable errors:} Physically, one
of the most important facts about quantum error-correction is that if
${\cal R}$ is a recovery operation for a quantum operation ${\cal E}$
with operation elements $E_k$, then ${\cal R}$ also acts as a recovery
operation for any quantum operation ${\cal F}$ whose operation
elements $F_l$ can be expressed as linear combinations of the $E_k$.
It is this fact which allows us to focus attention on correcting a
discrete set of errors (usually the Pauli $I, X, Y$ and $Z$ errors)
since an arbitrary operation element on a qubit may be expressed as a
linear combination of those errors.

The analogous fact is also true for operator quantum error-correction.
Suppose ${\cal R}$ is a recovery operation for ${\cal E}$, with
respect to the decomposition $V = (A \otimes B) \oplus C$.  As noted
in the proof that \textbf{[a]} implies \textbf{[b]}, we have $R_j E_k
P = I \otimes N_{jk}$ for some set of operators $N_{jk}$ on $B$.
Suppose ${\cal F}$ is some other quantum operation whose operation
elements $F_l$ may be expressed as linear combinations of the $E_k$,
i.e., $F_l = \sum_k e_{lk} E_k$, where the $e_{lk}$ are complex
numbers.  Then it follows that $R_j F_l P = I \otimes \tilde N_{jl}$,
where $N_{jl} \equiv \sum_k e_{lk} N_{jk}$.  A direct computation
shows that ${\cal R}$ also acts as a recovery operation for ${\cal
  F}$, which concludes the proof. \textbf{QED}

\textbf{Generalizations of operator quantum error-correction?}  We
have studied the storage of quantum information in a subsystem $A$ of
a subspace of $V = (A \otimes B) \oplus C$. Is it possible to store
quantum information in some other way within $V$?  For example,
perhaps it is possible to decompose $A$ into two subspaces, $A = A_1
\oplus A_2$, and store information solely in $A_1$.  However, if we do
this then the total vector space may be decomposed as $V = (A_1
\otimes B) \oplus \tilde C$, where $\tilde C = (A_2 \otimes B) \oplus
C$, and thus this is a special case of the type of decomposition
already considered.  More generally, the distributive properties of
the tensor product and direct sum ensure that no matter how we try to
``nest'' information within multiple layers of subspaces and
subsystems, the end result can always be expressed as a decomposition
of the form $V = (A \otimes B) \oplus C$, where the subsystem $A$ is
used to store the quantum information.

\textbf{Information-theoretic characterization of correctability:} For
quantum error-correcting codes an information-theoretic necessary and
sufficient condition for the correctability of trace-preserving ${\cal
  E}$ was found in~\cite{Schumacher96b}, and subsequently generalized
to non-trace-preserving ${\cal E}$ in~\cite{Nielsen98a}.  We now find
a set of information-theoretic necessary and sufficient conditions for
operator quantum error-correction, generalizing the earlier
conditions, and actually simplifying those in~\cite{Nielsen98a}.

Most of the work has already been done in arriving at condition
\textbf{[c]}, above.  Suppose we normalize the state $|\psi'\rangle$
so $\rho'$ and the corresponding reduced density matrices all have
trace 1.  The subadditivity inequality for entropy (see p.~515 and~516
of~\cite{Nielsen00a}) implies that $S(\rho_{R_A R_B E}') \leq
S(\rho_{R_A}')+S(\rho_{R_B E}')$, with equality if and only if
$\rho_{R_A R_B E}' = \rho_{R_A}' \otimes \rho_{R_B E}'$.  It follows
that a necessary and sufficient condition for ${\cal E}$ to be
correctable is that $S(\rho_{R_A}') + S(\rho_{R_BE}') = S(\rho_{R_A
  R_B E}'$.  This may be rewritten in a more convenient form by noting
that $S(\rho_{R_A}') = S(\rho_{R_A}) = S(\rho_A)$, and that
$S(\rho_{R_AR_BE}') = S(\rho_{V}')$.  This gives us the following
necessary and sufficient condition for ${\cal E}$ to be correctable.
(Note that in an obvious notation $S(\rho_A) = \log(d_A)$, where $d_A$
is the dimension of system $A$, since $A$ is initially maximally
entangled with $R_A$.)

\textbf{[d]}: $S(\rho_A) = S(\rho_{V}')-S(\rho_{R_B E}')$. 

The conditions~\textbf{[d]} generalize the necessary and sufficient
conditions in~\cite{Schumacher96b,Nielsen98a}
(c.f.~\cite{Schumacher02a,Ogawa05a}), which correspond to the case
when $B$ is trivial.  Note that~\cite{Schumacher96b,Nielsen98a} allow
$A$ and $R_A$ to start out in a state which is not maximally
entangled, but rather are merely of full Schmidt rank.  Our arguments
are easily generalized to this case.

\textbf{Data processing inequality:} We have described the condition
\textbf{[d]} as information-theoretic, but have not suggested an
information-theoretic interpretation of the quantities involved.  Such
an interpretation is suggested by the following argument, which
generalizes the coherent information introduced
in~\cite{Schumacher96b}.  \cite{Schumacher96b} showed that the
coherent information satisfied a monotonicity property known as the
\emph{quantum data processing inquality}, which states that quantum
information can only ever be lost as it is passed through multiple
quantum channels; once lost, quantum information can never be
recovered.  The coherent information and quantum data processing
inequality played a key role in subsequent investigations of the
quantum channel
capacity~\cite{Schumacher96a,Schumacher96b,Barnum98a,Barnum98b,Lloyd97a,Shor02a,Devetak05a}.

We now prove an analogue of the quantum data processing inequality
which applies to operator quantum error-correction.  Our analysis is
based on the conditional entropy of $R_A$ given $V$, $-S(R_A|V) \equiv
S(V)-S(R_A V)$, which generalizes the coherent information.  The
following argument suggests that this may be regarded as a measure of
the amount of quantum information about the initial state of $A$ which
is still stored in $V$.  Suppose we apply a sequence of
trace-preserving quantum operations ${\cal E}_1,{\cal E}_2, \ldots$ to
$V$.  Standard monotonicity properties of the conditional entropy
imply that
\begin{eqnarray} \label{eq:gen_data_processing_inequality}
  -S(R_A|V) \geq -S(R_A'|V') \geq -S(R_A''|V'') \geq \ldots,
\end{eqnarray}
where a single prime indicates that ${\cal E}_1$ has been applied, a
double prime indicates that ${\cal E}_2 \circ {\cal E}_1$ has been
applied, and so on.  Eq.~(\ref{eq:gen_data_processing_inequality}) is
a generalization of the data processing inequality obtained
in~\cite{Schumacher96b}.

Condition~\textbf{[d]} is easily seen to be equivalent to the
condition $-S(R_A'|V') = -S(R_A|V)$, i.e., that the coherent
information be preserved by the operation ${\cal E}$.

Indeed, a consequence of~(\ref{eq:gen_data_processing_inequality}) is
an informative alternative proof of the necessity of \textbf{[d]}.
Suppose ${\cal E}_1 = {\cal E}$ and ${\cal E}_2 = {\cal R}$.  The fact
that ${\cal R}$ restores the information stored in $A$ implies that
$-S(R_A|V) = -S(R_A''|V'')$.  It follows
from~(\ref{eq:gen_data_processing_inequality}) that we must have
$-S(R_A'|V') = -S(R_A|V)$, which implies~\textbf{[d]}.


\textbf{Conclusion:} Operator quantum error-correction is a recently
introduced technique for stabilizing quantum information, which
generalizes and unifies previous approaches, including standard
quantum error-correcting codes, decoherence-free subspaces, and
noiseless subsystems.  In this paper we've developed algebraic and
information-theoretic necessary and sufficient conditions for operator
quantum error-correction, and used these conditions to develop an
elegant representation theorem for a wide class of correctable noise
processes, as well as generalizations of the coherent information and
quantum data processing inequality.  Open problems include the
systematic investigation of specific operator quantum codes, and the
investigation of techniques for fault-tolerant quantum information
processing using operator quantum codes.

\acknowledgments

Thanks to Dave Bacon, Steve Bartlett, Dominic Berry, Jennifer Dodd,
and Andrew Doherty for helpful discussions and suggestions for
improvement.


\end{document}